\documentclass[aps,preprint,a4paper,showpacs,amsmath,amssymb,floatfix,prb]{revtex4}
\usepackage{graphicx}
\DeclareGraphicsExtensions{.pdf}
\newcommand{\Op}[1]{{\bf {\hat {#1}}}}

\begin{document}

\title{Quantifying the Unitary Generation of Coherence From Thermal Quantum Systems} 

\author{S.Kallush$^{1,2}$, A. Aroch$^{2}$, R. Kosloff$^{2}$}
\affiliation{$^1$  Department of Physics and Optical Engineering, ORT-Braude College, P.O. Box 78, 21982 Karmiel, Israel
\\$^2$ The Fritz Haber Research Center, The Hebrew University of Jerusalem,
Jerusalem 91904, Israel}

\begin{abstract}  
The unitary generation of coherence from an incoherent thermal state is investigated.
We consider a completely controllable Hamiltonian allowing to generate all possible unitary transformations.
Optimizing the unitary control to achieve maximum coherence
leads to a micro-canonical energy distribution on the diagonal energy representation.
We demonstrate such a control scenario starting from a Hamiltonian utilizing optimal control theory for unitary targets.
Generating coherence from an incoherent initial state always costs external work.
By constraining the amount of work invested by the control, maximum coherence leads to a canonical energy population distribution. When the optimization procedure constrains the final energy too tightly local suboptimal traps are found. 
The global optimum is obtained when a small Lagrange multiplier is employed to constrain the final energy.
Finally, we explore constraining the generated coherence to be close to the diagonal in the energy representation.
\end{abstract}

\maketitle


\section{Introduction}
Coherence is associated with transient quantum states, in contrast  
equilibrium thermal quantum systems have no coherence. A signature of a stationary state is that it commutes with
the Hamiltonian $[\Op H , \Op \rho_{st} ]=0$. In addition, a thermal state is also a passive one, characterized by a monotonic decreasing probability distribution of the occupation of its energy levels \cite{haag1974,lenard1978,pusz1978passive}. The issue addressed in this study is the optimal
generation of coherence from a thermal state and its minimal cost in energy.  

Generating coherence from a passive state involves a cost in work. Coherence generation
requires energy excitations.
The other extreme is adiabatic evolution which will maintain a passive energy state with no coherence. 
The tradeoff between work and coherence has been noticed \cite{kieu04,plastina2014,deffner2010generalized}. In this study we want to quantify the work cost\cite{misra2016energy}.

A density operator of a thermal state $\Op \rho_T = \frac{1}{Z}e^{-\beta \Op H}$ is diagonal in the energy representation,
where $\Op H$ is the Hamiltonian of the system, $\beta =\frac{1}{k_B T}$, and $Z$ is the system's partition function: 
$Z= \textrm{Tr} \{ e^{-\beta \Op H}\}$. Coherence is associated with off-diagonal elements of the density operator.
Quantifying coherence inecessitates the determination of dynamical field-free evolution of non-diagonal elements of the density matrix. 
Examples for such observables are molecular alignment and orientation\cite{felker,damari16}.

There have been several suggestions to quantify the amount of generated coherence \cite{k108,baumgratz2014quantifying}. 
In this study we will use a distance metric ${\cal D}$ and ${\cal C}$ (see below).

\section{Maximum Coherences, Shannon Entropy, and the Micro-canonical ensemble}

We study the creation of coherence from a passive state by a unitary evolution
generated by the control Hamiltonian ${\Op H}$:
\begin{equation}
\Op{H}(t) = \Op H_0 + \Op \mu \cdot f(t)
\label{eq:chamil}
\end{equation}
where $\Op H_0$ is the stationary Hamiltonian, $\Op \mu$ is the control operator, $f(t)$ is the time dependent control field. The dynamics generated by the control Hamiltonian are given by the unitary evolution operator:
\begin{equation}
i \hbar \frac{d}{dt} \Op U_c = \Op H(t) \Op U_c~~~~~~\Op U_c(0) =\Op I .
\label{eq:evolution}
\end{equation}

We assume complete controllability meaning that the control field can generate any unitary transformation $\Op U_c$ in the Hilbert space of the system. We seek a unitary $\Op U_c$ which will transform an initial thermal state $\Op \rho_T$
to a final state $\Op \rho_f$ with maximum coherence
\begin{equation}
\Op {\rho}_f = \Op {U}_c^\dagger {\Op \rho_T}\Op {U}_c
\label{eq:rhoif}
\end{equation}  

We now need to quantify the coherence. 

The Shanon entropy \cite{shanon48} associated with a complete measurement of the observable $\langle \Op A \rangle$ is
\begin{equation}
{\cal S}_A = - \sum_j p_j \ln p_j
\label{eq:shanon}
\end{equation}
where $p_j= \langle \Op P_j \rangle = \textrm{Tr}\{\Op \rho \Op P_j\}$ is the probability of outcome $j$ considering the spectral decomposition
of the operator $\Op A = \sum \alpha_j \Op P_j$. A special case is the energy entropy ${\cal S}_E$
associated with a complete measurement of $\Op H$.  
In information theory terms, the entropy with respect to a variable quantifies 
the amount of classical information obtained by a complete
measurement. If the evolution operator $\Op U$ does not commute with $\Op A$ 
the entropy of the observable ${\cal S}_A$ will change.  

An invariant to the evolution is the von Neumann entropy ${\cal S}_{vN}$ \cite{vNeumann}:
\begin{equation}
{\cal S}_{vN} = -  {\textrm{Tr}\left( {\hat \rho \ln \hat \rho } \right)} 
\end{equation}
Due to its invariance with respect to unitary transformations $S_{vN}$ 
is often used to quantify the total information content of quantum systems \cite{nielsen2000quantum}. 
As shown in\cite{k282} $S_{vN}$ is the minimal entropy that could be obtained by a complete measurement
of all possible operators. Therefore, ${\cal S}_{vN} \le {\cal S}_A$. For a thermal state $\Op \rho_T$ the energy entropy
and the von Neumann entropies coalesce ${\cal S}_E ={\cal S}_{vN}$.
 
 A quantifier of distance between two states $\Op \rho_a$ and $\Op \rho_b$ is the divergence \cite{lindblad74}:
 \begin{equation}
{\cal D} (\Op \rho_a | \Op \rho_b )   = \textrm{Tr} \{ \Op \rho_a \ln \Op \rho_a - \Op \rho_a \ln \Op \rho_b \}
\end{equation}
${\cal D} (\Op \rho_a | \Op \rho_b ) \ge 0$ with equality when $\Op \rho_a =\Op \rho_b$. 
A measure of coherence is the distance between a state $\Op \rho$ and $\Op \rho_E$ 
where $\Op \rho_E$ is the diagonal part of the density operator $\Op \rho$ in the energy representation.
We can quantify this distance by ${\cal D}_E (\Op \rho | \Op \rho_E )  $ so it becomes the difference
between the von Neumann entropy and the energy entropy \cite{k311,uzdin2018global}:
\begin{equation}
{\cal D}(\Op \rho| \Op \rho_E) = {\cal S}_{E}-{\cal S}_{vN} 
\end{equation}
For a thermal state $\Op \rho_T$ the divergence vanishes. Under unitary transformation ${\cal S}_{vN}$ is conserved but the energy entropy
${\cal S}_E$ can increase and with it the divergence ${\cal D}(\Op \rho| \Op \rho_E) $, therefore
maximizing ${\cal S}_{E}$ will lead to maximum  coherences.

An alternative direct measure for coherences is:
\begin{equation}
{\cal C} = \frac{2N}{N-2}\sum\limits_{i,j > i} {{{\left| {{\rho _{ij}}} \right|}^2}}
\label{codef} 
\end{equation}
where the sum of the magnitude of the off diagonal elements of ${\Op \rho}$ in the energy representation is measured. $\cal{C}$ is normalized such that $C = 1$ is obtained  for a fully coherent pure state.

We start by analyzing the maximum generated coherences in an unconstrained case. 
The entropy is a monotonic function of the temperature. It is zero for $T = 0$, and the thermal state with the maximum (either von Neuman and Shannon) entropy is the microcanonical ensemble in which $p_j = 1/N$, and  ${\cal S}_{vN} = {\cal S}_E = \ln N$. This corresponds to a state with effective $T \to \infty$. Now, any initial thermal state with lower $T$ will have lower ${\cal S}_E$. Therefore the unitary transformation that will maximize 
${\cal D}(\Op \rho| \Op \rho_E)$ is the one that transforms ${\Op \rho}_T$ into a population distribution similar to the microcanonical ensemble on the energy diagonal.
The absolute maximal generated coherences for any system  is therefore starting from 
an initial pure state, which for a thermal state with $T=0$ is the ground state and vanishing entropy ${\cal S}_{vN} = 0$. 

\subsection*{Computational Control Demonstration and Model}

The control scenario is demonstrated with the model of a many body single mode Bose-Hubbard 
double well system\cite{mahan2013many}.  This many-body Hamiltonian for $n$ atoms  is equivalent to  a
system of $N = 2j+1$ sub-levels with angular momentum $j=n/2$ \cite{k286}. The Hamiltonian model becomes:
\begin{equation}
\Op {H}_0 = {U}\Op{J}_z^2 + \Delta \Op{J}_x
\label{hmodel}
\end{equation}
Where $\Op{J}_i$ are the projections of the total angular momentum on the $i$ axis. 
This Hamiltonian is completely controllable, for more details, see ref \cite{k286}.  
To enable systematic comparison between systems with different sizes, the energy of the system is normalized so the effective free Hamiltonian of the system is taken as $\Op H_n = {{\Op H}_0} /\textrm{Tr}({\Op H_0}^2)$.  

\begin{figure}
    \includegraphics[scale=0.4]{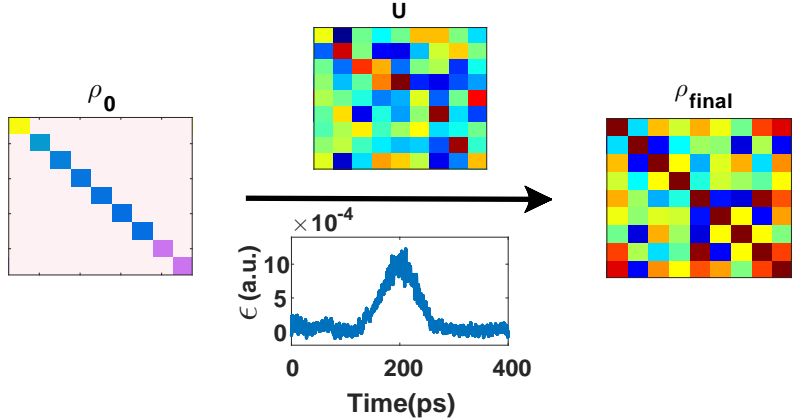}
    \caption{Unitary transformation generating maximum coherence: Results of the Optimal-Control-Theory field that was generated to take the thermal state with $j = 4$ ($N=9$) and $\beta_0=0.2$ with the Hamiltonian of eq. (\ref{hmodel}) into the micro-canonical population distribution. Left panel: initial thermal density matrix. Uper and lower panels: absolute values of the obtained unitary transformation, and its generating field, respectively. Right panel: absolute values for the transformed density matrix.}
    \label{fig0}
\end{figure}

Complete controllability \cite{k193}, means that optimal control theory  will lead 
to  the field that generates the desired unitary transformation. 
To illustrate this ability on the current context, we use the Hamiltonian of Eq. (\ref{hmodel}), 
with the initial thermal state with inverse temperature $\beta_0 = 1/k_b T_0 = 0.2 $ $1/Hartree$. 
The control hamiltonian $\Op H = \Op H_n+ \epsilon(t){\Op J_z}$ was then used
to generate the target unitary transformation. 
This unitary operator transforms the thermal state into the micro-canonical distribution
on the diagonal. By construction it leads to maximum coherence.
Figure \ref{fig0} presents an image matrix-plot for the initial and the final absolute values of the density matrices, 
the computed unitary transformation, and the field $\epsilon(t)$ that produced it.  

The optimality of the micro-canonical probability distribution with respect to coherences, is demonstrated by
the following numerical optimization:
Starting with given ${\Op \rho}_T$ with corresponding $\beta_0 = 1/k_{b}T_0$, find the unitary transformation: 
$\Op U^{op} = \exp(i \Op V^{op})$ such that the functional  ${\cal D}(\Op \rho| \Op \rho_E)$ is maximum. 
For an $N$ level system, this defines a control problem with dimensionality $N \times N$ free control parameters, 
defined by the hermitian matrix ${\Op V}^{op}$.

\begin{figure}
    \includegraphics[scale=0.6]{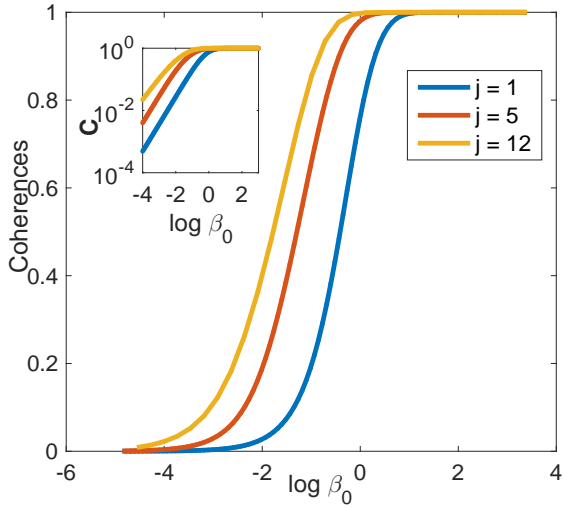}
    \caption{Micro-canonical ensemble: Maximal coherences defined by eq.\ref{codef}, as a function of the initial verse temperature $\beta$, for different sizes of the system. Inset: A log-log plot of the main panel.}
    \label{fig1}
\end{figure}

Figure \ref{fig1} presents the value of the direct optimal total coherences measure ${\cal C}$ defined  in Eq. (\ref{codef}), 
as a function of the logarithm of the inverse temperature $\beta_0$ for different sizes of the system. 
For all  cases, the optimal population distribution was found to be the microcanonical ensemble. 
As expected, for increasing temperature, the level of coherence generated decrease smoothly so that at $T \to \infty$ no coherence is generated, vanishing exponentially with the initial 
$\beta_0$ (see left inset). For larger systems, coherence is generated for higher initial  temperatures.  
Note that the optimum is highly degenerate, represented by a $N(N-1)/2$-fold  dimensional sphere with a vanishing radius for $T \to 0$ and $T \to \infty$, and a maximum in an intermediate $\beta$. Being loosely constrained and highly degenerate, the optimization is found to be globally trap-less. 
That is, for all the sizes of the system and for any initial guesses for the unitary transformation, no suboptimal solutions were found \cite{chakrabarti2007quantum,rabitz2005landscape}. 
All the solutions converged to the global optimum of the micro-canonical distribution $p_i = 1/N$. Difficulties to converge to the optimum were found for $j > 12$ ($N=25$).
The extremely large dimensionality of the optimization problem (above $625$ optimization parameters) 
cause difficulty in computing the numerical gradient.
  
\section{Energy constraint and the Canonical probability distribution}

The micro-canonical probability distribution on the diagonal is the state with utmost coherences for a $N$ level system. 
This state, however, corresponds to a uniform population distribution, which effectively matches infinite temperatures for the probability distribution. 
The natural constraint in addition is  a target with finite energy. 

The state with maximum entropy constrained by an average energy $\langle \Op H \rangle =  E$ is the thermal state $\Op \rho_T$ and its entropy ${\cal S}_{vN} = {\cal S}_{E} = \ln Z + \beta E$ \cite{katz1967principles}. 
For $T \rightarrow \infty$ the thermal state converges to the micro-canonical state $\Op \rho_T \rightarrow \Op \rho_{mc}$.
Therefore, a unitary transformation with a target state where the populations are canonically distributed will maximize  coherences for a predetermined final energy. Consequentially, for any initial thermal incoherent state, generating coherences requires an investment
of energy. The process is reversible, nevertheless returning to the original energy will necessarily  erase  the coherences. 
Moreover,  for any passive states, this also means that the energy of a thermal state cannot be reduced by a unitary transformations due to the fact that coherences cannot be produced.

\subsection*{Numerical example}

The constraint on the final energy is introduced by a Lagrange multiplier  in addition to the previous optimization of the  divergence.
The optimization problem is now in the following form: Starting with with the inverse temperature $\beta_0$ and ${\Op \rho}_T$, find the optimal unitary transformation: $\Op U^{op} = \exp(i \Op V^{op})$ such that $J= {\cal S}_{E} - \lambda \left|E-E_f \right|$ is maximal. $\lambda$ is the Lagrange multiplier that imposes the energy constraint.  

\begin{figure}
    \includegraphics[scale=0.6]{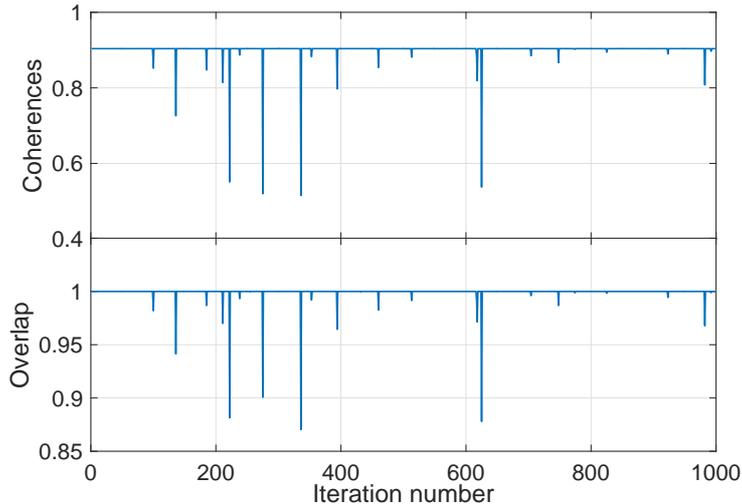}
    \caption{Canonical ensemble. Upper panel: The values for ${\cal C}$ for unsorted $1000$ optimization runs. The parameters for the runs were taken as $j = 1$, the initial temperature is $\beta = 3$ and the final is $\beta_f = 0.3$. Lower panel: Overlap between the final populations of the density matrix and the target thermal state. $\lambda=0.3 = \lambda_0$ (see text) is taken.}
    \label{fig2}
\end{figure}

The optimization with the additional constraint leads to traps: many local  suboptimal solutions. 
We use this feature to corroborate the result of this section. Figure \ref{fig2} presents in the upper panel the measure of 
the coherence ${\cal C}$ for $1000$ optimization runs with different initial random guesses for the unitary transformation. 
At the bottom panel the overlap between the  diagonal part of the final density matrix $p_i^f$ and the thermal population at the target energy 
$p_i^T$, defined by: $O = \sum\limits_i {\sqrt {p_i^T p_i^f} } $, is shown. 

The suboptimal traps are clearly visible. Moreover, the correspondence between the traps in the plots is strict, and, as expected a complete overlap with the target thermal distribution leads also to a maximum in the coherences. 

\begin{figure}
    \includegraphics[scale=0.6]{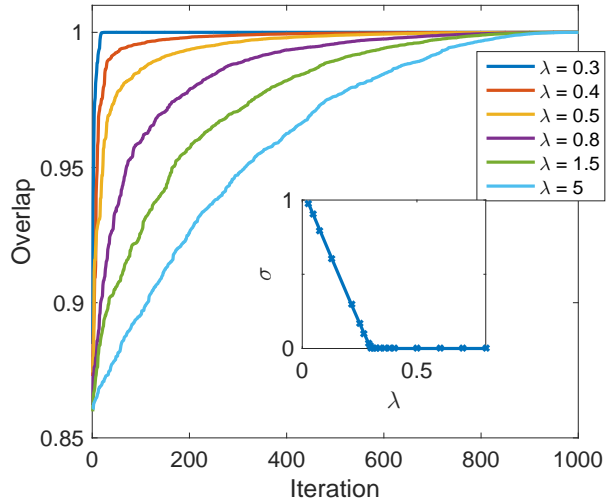}
    \caption{$\lambda$ and number of traps. Main panel: Sorted overlap (see main panel of fig.\ref{fig2}) for $1000$ runs, for different values of the Lagrange multiplier values $\lambda$. 
    Inset: the mean error in the resulted energy as a function of the Lagrange multiplier $\lambda$.}
    \label{fig3}
\end{figure}

Remarking on the number of traps in the system: 
The inset in figure \ref{fig3} shows the averaged relative error in the final energy 
of the state $\sigma$  as a function of the Lagrange multiplier $\lambda$:
\begin{equation}
\sigma = \frac{E-E_T}{E_T}
\end{equation}
where $E$ is the actual energy of the final state, and $E_T$ is the target thermal energy. For very small $\lambda \to 0$, the energy constraint is too weak to impact the optimization, and the solution does not converge to the correct target energy. 
However,  violation of the energy constraint decreases linearly with $\lambda$, 
probably due to the linearity of the energy in the constraint. 
Adequate numerical tolerance was reached for some $\lambda_0$. 
The main panel of the figure show the sorted values for the overlap between 
the final density matrix and the thermal target, for $1000$ optimization runs, and different values of $\lambda$. 
For the case presented here $\lambda_0 = 0.3$, it is interesting to see that the number of traps is minimal very close to $\lambda_0$, and increases monotonically with the increase of the Lagrange multiplier. This is a result of the relative weight of the optimized quantity ${\cal D}$ which is relatively decreased, resulting in suboptimal overlap and coherences.     

\section{Additional constraints imposing coherence predominately close to the diagonal}

Typically in controllable problems the control operator is biased to couple adjacent energy levels.
For example the electric dipole or the polarizability tensor operators\cite{seidman}. 
In the context of molecular spatial directions for example, both operators are related 
to $cos \theta$ and $cos^2\theta$, where $\theta$ is the angle between the molecular and spatial axis. 
The two kinds of interactions lead to molecular orientation and alignment, respectively. 
These light-matter coupling operators connect directly only adjacent (or next-to-adjacent) $j$ levels. 
Hence, coherences will be first generated at near $j$ proximity. Higher order of coupling, 
between distant energy levels, are not forbidden but they are harder to achieve. In control terminology, 
the problem is still controllable, but is not invertible, i.e., the solution for the control problem will be very hard or even unfeasible.  

\begin{figure}
    \includegraphics[scale=0.5]{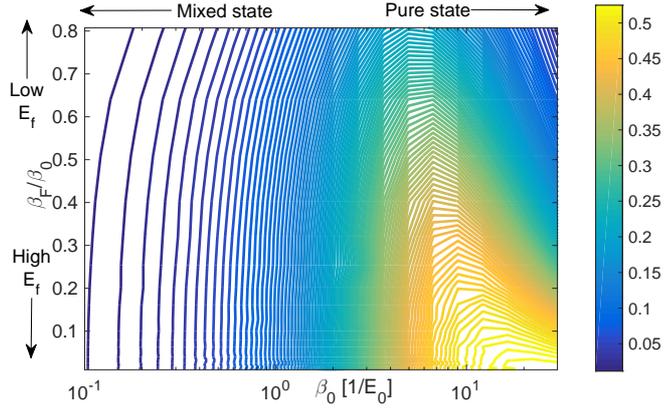}
    \caption{Coherences and phase space proximity I. Maximal values for the generalized coherences operator $\Op O$, defined in eq. \ref{O_def}, as a function of the initial temperature $\beta_0$ and final effective temperature $\beta_F$. The other parameters taken here are $\alpha = 0.04$, and $j=3$.}
    \label{fig4}
\end{figure}

To study  the development of coherences, the optimization is modified,with the target being
\begin{equation}
\hat O = \exp \left( {\alpha \hat \mu } \right) - diag\left( {\exp \left( {\alpha \hat \mu } \right)} \right)
\label{O_def}
\end{equation}
where
\begin{equation}
\hat \mu  = \left( {\begin{array}{*{20}{c}}
0&1&{}&{}&{}\\
1&0&1&{}&{}\\
{}&1&{...}&{...}&{}\\
{}&{}&{...}&{}&1\\
{}&{}&{}&1&0
\end{array}} \right) 
\end{equation}
is a simplified dipole  operator. Under these definitions, $\left<{\Op O}\right>$ is a measure of the amplitude of the 
off-diagonal coherences. 
For small values of $\alpha$, $\Op{O} \to \Op{\mu}$  only adjacent coherences are emphasized, and for increasing $\alpha$ 
the whole matrix is covered,  merging into the micro canonical ensemble result of the previous section. 
Figure \ref{fig4} displays a contour plot of the optimal expectation value of $\Op{O}$ under the unitary transformation 
that gives optimal $\left<{\Op O}\right>$ for the system with $j=3$ ($N=7$) and $\alpha = 0.04$. 
The values are plotted as a function of the initial temperature $\beta_0$, and the final energy 
defined by the effective temperature $\beta_F$. The $x$-axis values are an  indication of the initial purity of the system, 
so that the system is pure for $\beta_0 \to \infty$, and highly mixed for vanishing $\beta_0$. 
The $y$-axis is an indication for the resulted energy of the system. 
The optimization problem in this case contains, of course, many traps, 
and the globally optimal results displayed in this section were verified by initiating the optimization with $10000$ 
random guesses for the transformation, and choosing the (converged) maximum value that was obtained for $\Op O$. 
The maximum of the coherence generated  is obtained for an initial pure state that is transformed to final infinite 
temperature and distribution on the diagonal, i.e., the micro-canonical ensemble. 

\begin{figure}
    \includegraphics[scale=0.5]{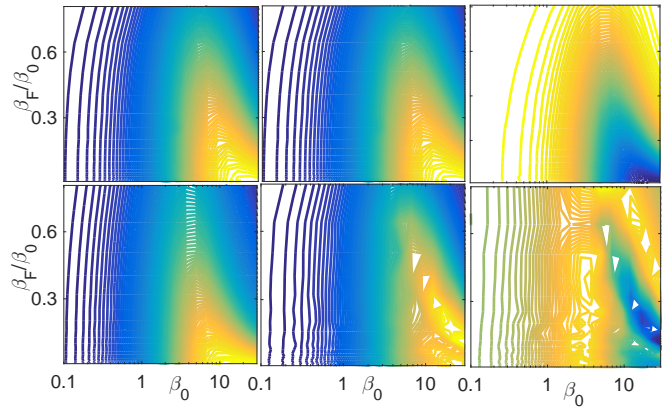}
    \caption{Coherences and phase space proximity II. Upper panels: Similar to fig. \ref{fig4}, plotted in the upper left panel reference. The middle and right upper panels are the expectation values of $\Op \mu$ and ${\Op\mu}^2$ at the optimal conditions that were obtained for the left upper panel of the operator $\Op O$. The rest of the parameters are similar to fig \ref{fig4}. Lower panels: Same as the upper panels, with $\alpha = 40$.}
    \label{fig5}
\end{figure}

Figure \ref{fig5} expands the result of figure \ref{fig4} to other regimes and operators. 
As a reference, the results of fig \ref{fig4} are displayed on the upper left panel of the figure. 
The middle upper panel displays the  expectation values of $\Op \mu$ 
of the target state which optimizes $\Op O$.
For this case $\alpha =0.04 \ll 1$ the optimization of the two operators converges to the same outcome. 
The right upper panel of figure shows the same dependency, under the same conditions, 
for the non-diagonal part of the operator $\Op{\mu}^2$, which quantifies
the second row off-diagonal elements. It is interesting to note that the maximum conditions for coherences for $\Op \mu$ 
is associated with minimal coherences for $\Op{\mu}^2$.
This feature of inverted extremum seems to hold also for higher orders of $\Op \mu$.
The lower panels of fig. \ref{fig5} present the same veriables of the upper panel, but here 
$\alpha= 40 \gg 1$. For this case, the maximum of $\Op O$ is still found for an initial pure state and maximum final energy, 
but the entire structure does not overlap with the conditions for the linear dipole coherences. 
The structure that flips the signs of the minimum and maximum with the order of $\Op \mu$ is nevertheless maintained. 
As a final remark, it is worth mentioning that the same procedure was tested also for a generator of the form $\Op{O}_2 \propto \exp(\alpha \mu^2)$. For this generator the optimal transformations for $\Op O_2$ were found to contain only coherences of even order in $\Op \mu$, and the inversion of the minimum and maximum takes place now with $\Op \mu^{2}$.

\section{Summary}

The existence of coherence in a system is a signature of its quantum properties. In this work the ability to induce coherence into systems by means of an external control field was investigated. We found that for any thermal state the coherence generated will 
become maximal when the system is transformed into a population distribution that matches the microcanonical energy distribution.
Adding an energy constraint to the final target state leads to an optimal population distribution that matches the canonical distribution. 
Note that the maximum coherence could be also obtained indirectly by imposing a complete {\it population transformation} 
target\cite{k323} on the optimization procedure instead of the full unitary transformation. 

 Finally, it is worth while to mention that the insight  obtained from the control problem 
is simple and robust, nevertheless their physical realization is non trivial. 
Even where the system  is controllable, the actual typical operators that generate 
the transformation have low connectivity within the whole Hilbert space. 
In that case the it is difficult to generate coherence between distant energy states. 
The last section  was devoted to this restriction, and its interesting properties and symmetries  were discussed.
The present study is a link between optimal control theory and quantum thermodynamics \cite{k281} pointing to the work cost
required to generate coherence.

\section{Acknowledgement}
This work was supported by the US-Israel Binational Science Foundation through Grant No. 2012021. This research
was also supported by the Israel Science Foundation (Grant No.
510/17).


\end{document}